\documentclass[osajnl,twocolumn,showpacs,superscriptaddress,10pt]{revtex4-1} 
\usepackage{amsmath,amssymb,graphicx}

\usepackage{mathtools}
\usepackage{xfrac}

\begin{document}

\title{Microresonator-based high-resolution gas spectroscopy}

\author{Mengjie Yu}\email{Corresponding author: my2473@columbia.edu}
\affiliation{Department of Applied Physics and Applied Mathematics, Columbia University, New York, NY 10027}
\affiliation{School of Applied and Engineering Physics, Cornell University, Ithaca, NY 14853}

\author{Yoshitomo Okawachi}
\affiliation{Department of Applied Physics and Applied Mathematics, Columbia University, New York, NY 10027}

\author{Austin G. Griffith}
\affiliation{School of Applied and Engineering Physics, Cornell University, Ithaca, NY 14853}

\author{Michal Lipson}
\affiliation{Department of Electrical Engineering, Columbia University, New York, NY 10027}

\author{Alexander L. Gaeta}
\affiliation{Department of Applied Physics and Applied Mathematics, Columbia University, New York, NY 10027}

\begin{abstract}In recent years, microresonator-based optical frequency combs have created up opportunities for developing a spectroscopy laboratory on a chip due to its broadband emission and high comb power. However, with mode spacings typically in the range of 10 - 1000 GHz, the realization of a chip-based high-resolution spectrometer suitable for gas-phase spectroscopy has proven to be difficult.  Here, we show mode-hop-free tuning of a microresonator-based frequency comb over 16 GHz by simultaneously tuning both the pump laser and the cavity resonance. We illustrate the power of this scanning technique by demonstrating gas-phase molecular fingerprinting of acetylene with a high-spectral-resolution of $<$ 80 MHz over a 45-THz optical bandwidth in the mid-IR. Our technique represents a significant step towards on-chip gas sensing with an ultimate spectral resolution given by the comb linewidth.
\end{abstract}


\maketitle 


Mid-infrared (Mid-IR) frequency combs\cite{schliesser_mid-infrared_2012} can enable broadband gas-phase spectroscopy\cite{vainio_mid-infrared_2016, cossel_gas-phase_2017} with unprecedented measurement capabilities and new applications in medical diagnostics,  astrochemistry, atmospheric monitoring, remote sensing and industry control. An optical frequency comb (OFC)\cite{udem_optical_2002, cundiff_textitcolloquium_2003} has a broad optical spectrum consisting of evenly spaced, spectrally sharp lines. The frequency of each comb line can be written as $f_{m}$ = $f_{ceo}$ + $mf_{r}$, where $f_{ceo}$ is the carrier-envelope offset frequency and $f_{r}$ is the repetition frequency. Compared to tunable lasers, such broadband combs enable an accurate measurement of a large number of molecular absorption lines and species simultaneously in a complex environment\cite{spaun_continuous_2016}. The most widely used OFC is based on modelocked femtosecond lasers\cite{udem_optical_2002, jones_carrier-envelope_2000} and has been previously utilized for high-resolution spectroscopy\cite{marian_united_2004, diddams_molecular_2007}. Recently, there have been interest in the development of OFC sources in the mid-infrared (mid-IR) fingerprinting regime where the gas phase has strong fundamental vibrational transitions. However, the extension of OFC's into the mid-IR has proven to be challenging. Modelocked lasers emitting beyond 2.5 $\mu$m have not yet been demonstrated\cite{schliesser_mid-infrared_2012}, and nonlinear optical techniques, such as difference frequency generation\cite{cruz_mid-infrared_2015, pupeza_high-power_2015, mayer_offset-free_2016} and optical parametric oscillation\cite{adler_phase-stabilized_2009, maidment_molecular_2016, smolski_coherence_2016, vainio_fully_2017}, are often used to convert an OFC from the near-IR to the mid-IR with additional system complexities. In addition, the repetition rate of the OFC needed for gas-phase spectroscopy is on the order of 100 MHz especially for probing Doppler-broadened absorption linewidths\cite{baumann_spectroscopy_2011, yan_mid-infrared_2016-1}. Therefore, detection techniques with high instrumental resolution are often required to resolve individual comb lines of a low repetition rate OFC. Comb-resolved molecular spectroscopy has been achieved using modelocked lasers in dual-comb systems\cite{baumann_spectroscopy_2011, millot_frequency-agile_2016, okubo_ultra-broadband_2015, yan_mid-infrared_2016-1} through long time averaging, high-resolution Fourier transform spectrometers (FTIR)\cite{maslowski_surpassing_2016} and VIPA-based spectrographs\cite{diddams_molecular_2007}. The generation of two OFC's with slightly different repetition rates in these systems typically require a complex setup, particularly in the mid-IR. The possibility of surpassing the resolution of FTIR and VIPA technique has been demonstrated by scanning the repetition rate of the OFC by changing the fiber cavity length\cite{maslowski_surpassing_2016, rutkowski_optical_2016, diddams_molecular_2007}. In the mid-IR, where direct laser sources and high quality optical components such as photodetectors (or arrays) remain in development, significant challenges exist for developing newer mid-IR OFC-based spectrometers with higher sensitivity and higher resolution in more compact, robust and less complex systems.

\begin{figure}[t]
\centering
\centerline{\includegraphics[width=7.0cm]{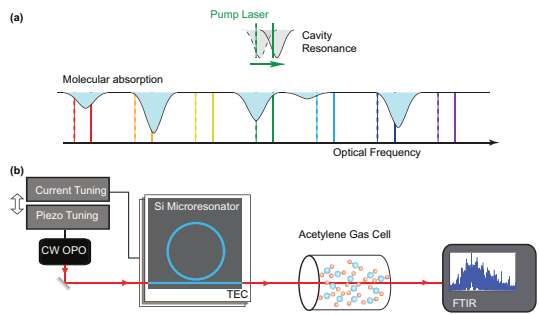}}
\caption{(a) Scheme for scanning comb spectroscopy. The generated comb lines follow the pump frequency tuning and probes different molecular transitions. (b) A silicon microresonator is pumped with a continuous-wave optical parametric oscillator. Mode-hop-free pump frequency tuning is achieved using a piezo controller, and the temperature of the chip is tuned using a current controller connected to a peltier underneath the chip, which thermally tunes the pump cavity resonance. The output from the chip is sent through a single-pass gas cell of acetylene and then recorded by a Fourier transform infrared spectrometer (FTIR).}
\label{Fig1}
\end{figure} 

Recently, chip-based mid-IR OFC's using quantum cascade lasers (QCL)\cite{hugi_mid-infrared_2012, villares_dual-comb_2014} and microresonators\cite{yu_silicon-chip-based_2016} have shown great promise as a new generation of more compact miniature spectrometers that allow for chip-scale integration. Over the past decade, microresonator-based OFC's \cite{delhaye_full_2008, delhaye_octave_2011, okawachi_octave-spanning_2011,wang_mid-infrared_2013, savchenkov_stabilization_2013, jung_electrical_2014, papp_microresonator_2014, huang_low-phase-noise_2015, griffith_silicon-chip_2015, xue_thermal_2016, joshi_thermally_2016, yu_mode-locked_2016, griffith_coherent_2016, yu_silicon-chip-based_2016,dutt_-chip_2016, suh_microresonator_2016, delhaye_phase-coherent_2016} have attracted tremendous interests since the platform enables highly compact devices, full integration, and powerful dispersion engineering which allows for broad optical bandwidths with moderate pump consumption\cite{okawachi_bandwidth_2014, yang_broadband_2016}. Several demonstration of direct-comb spectroscopy have been reported in the near-IR in silica\cite{suh_microresonator_2016}, silicon nitride\cite{pavlov_soliton_2017, dutt_-chip_2016} and fluoride microresonators\cite{wang_mid-infrared_2013} and recently achieved in the important molecular fingerprinting regime using silicon microresonators\cite{griffith_silicon-chip_2015, yu_silicon-chip-based_2016}. However, OFC's in such miniature devices inherently have large repetition rates typically from 10 to 1000 GHz, which preludes their use for high-spectral-resolution molecular spectroscopy. 

\begin{figure}[t]
\centering
\centerline{\includegraphics[width=7.0cm]{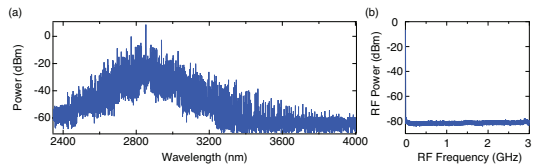}}
\caption{(a) The optical spectrum of a modelocked mid-IR frequency comb in a silicon microresonator. The full spectrum is collected using multiple bandpass filters before the FTIR and spans 2424 -- 3968 nm with a line spacing of 127 GHz. (b) The radio-frequency spectrum of the extracted free-carrier current indicates generation of a low-noise frequency comb.}
\label{Fig2}
\end{figure}

Here we report the first demonstration of a microresonator-based scanning OFC spectrometer suitable for gas-phase spectroscopy. We demonstrate mode-hop-free tuning of a modelocked mid-IR frequency comb in a silicon microreresonator over 16 GHz (0.53 cm$^{-1}$) via simultaneous tuning of temperature and pump laser frequency. The modelocked comb spans from 76 - 123 THz (1605 cm$^{-1}$) with a comb line spacing of 127 GHz (4.2 cm$^{-1}$). The molecular fingerprint of gas-phase acetylene is measured with a high-spectral resolution of $<$ 80 MHz (0.0026 cm$^{-1}$) despite using a low-resolution FTIR (15 GHz). This technique overcomes the resolution limitation induced by the ultra-small physical size of the integrated device and the instrumental lineshape and can find applications in precision spectroscopy\cite{liu_frequency-comb-assisted_2016}, tunable quantum sources\cite{ramelow_silicon-nitride_2015}, optical frequency synthesis \cite{huang_low-phase-noise_2015}, arbitrary waveform generation and optical coherence tomography. 

\begin{figure}[t]
\centering
\centerline{\includegraphics[width=7.5cm]{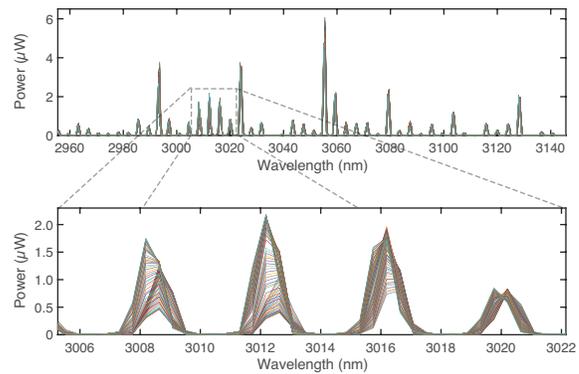}}
\caption{FTIR spectra of acetylene absorption measurement. A 2-cm-long single-pass gas cell gas cell is filled with 40 Torr of acetylene and 415 Torr of nitrogen. The OFC spectra is recorded at each 80 MHz of the pump shift with a low resolution of 15 GHz in the acetylene fingerprinting region (zoom-in: from 2960 to 3140 nm). The further zoom-in plot shows an optical range of 17 nm covering four comb lines where the first three comb lines are clearly overlapping with certain absorption features while the fourth one is not.}
\label{Fig3}
\end{figure}

Our approach is shown schematically in Fig. 1a. In a microresonator, the pump line is one mode of the OFC and has a fixed phase relationship with the rest of the generated comb lines. By joint frequency tuning of both the pump laser and its corresponding cavity resonance by one comb spacing, the OFC can be scanned to cover any spectral point within the comb bandwidth without any gaps. In addition, the spectral resolution for molecular spectroscopy can be reduced to the comb linewidth. In our case, the pump laser can be tuned over 100 GHz via piezo control and the cavity resonance is controlled via the thermo-optic effect\cite{delhaye_octave_2011, okawachi_octave-spanning_2011, jung_electrical_2014,xue_thermal_2016, joshi_thermally_2016}. The setup is shown in Fig. 1b. A high-$Q$ silicon microresonator with a 100-$\mu$m radius is dispersion engineered to have anomalous group-velocity dispersion beyond 3 $\mu$m for the fundamental TE mode, similar to Griffith, \emph{et al.}\cite{griffith_silicon-chip_2015}. The microresonator is pumped by a continuous-wave (CW) OPO ($<$100-kHz linewidth) at 2.85 $\mu$m (3508 cm$^{-1}$. The free carriers (FC) generated from three-photon absorption (3PA) are extracted with an integrated PIN junction and used to monitor the intracavity dynamics\cite{yu_mode-locked_2016, griffith_coherent_2016}. A thermoelectric cooler (TEC) is used to control the temperature of the silicon device, although future implementation can use an integrated heater on the chip to rapidly and precisely tune the temperature \cite{joshi_thermally_2016}. The output spectrum is measured using a commercial FTIR. As the CW pump laser is swept across the effective zero pump-cavity detuning, soliton modelocking is achieved as indicated by observation of the soliton steps\cite{herr_temporal_2014-1}, which results in a low-noise OFC with a narrow comb linewidth. In our experiment, we first generate a modelocked mid-IR OFC by tuning the pump laser into the cavity resonance at a pump power of 45 mW, and the spectrum consists of 378 comb lines with a spacing $f_{r}$ = 127 GHz and spanning 2.4 - 4.0 $\mu$m as shown in Fig. 2. We then tune both the pump frequency and the TEC current, which tunes the cavity resonance, while maintaining the same modelocked state. During the process, the DC component of the FC current is carefully kept the same in order to maintain a fixed relative pump-cavity detuning. This detuning can be used to switch between different soliton number states and has proven to be the key parameter of the modelocking dynamics\cite{guo_universal_2017}. We achieve a mode-hop-free tuning range of over 16 GHz which is 12.5$\%$ of $f_{r}$. This is equivalent to simultaneous mode-hop-free tuning of all 378 comb lines.  Ideally, a tuning range of one free-spectral-range (FSR) will enable continuous coverage over the entire spectral range. Currently, our tuning range is limited by the pump power variations over the larger bandwidth. However this is not a fundamental issue and a mode-hop-free tuning range over tens of GHz can be applied to a similar silicon microresonator of a larger radius to cover the generated spectral range. Furthermore, this scanning technique can be readily extended to other microresonator-based OFC's with integrated heaters\cite{joshi_thermally_2016}.

\begin{figure}[t]
\centering
\centerline{\includegraphics[width=7.5cm]{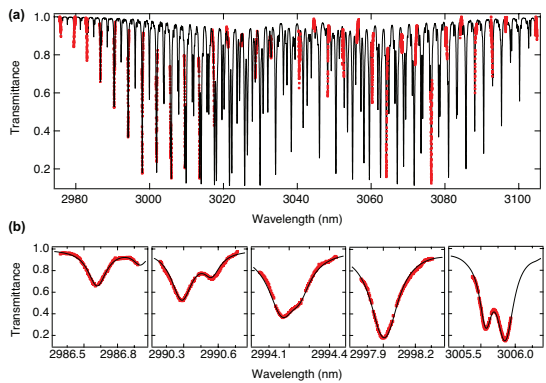}}
\caption{Molecular fingerprinting of acetylene. (a) The transmittance is calculated as the ratio between the spectrum with and without the gas cell. The frequency resolution in our measurement is 80 MHz. The transmittance spectrum is compared to the computed transmittance profile using the HITRAN database with Lorentzian lineshapes based on the density and buffer-gas pressure of the gas-cell. (b) Transmittance (zoom-in) of five measured absorption features at 2986.67 nm, 2990.39 nm,  2994.17 nm, 2998 nm, and 3005.76 nm, indicating our measurement clearly captures the detailed absorption lineshapes.}
\label{Fig4}
\end{figure}

We apply the scanning comb for an absorption measurement of gas-phase acetylene. The output is recorded with an FTIR after a 2-cm-long single-pass cell, which is filled with 40 Torr of acetylene and 415 Torr of nitrogen. Using this method, microresonator-based scanning OFC spectroscopy can break the resolution limitation of both the $f_{r}$ and the FTIR resolution. The spectral resolution of our system is limited by the comb linewidth in the modelocked state. Since our system is free running,  the resolution is limited primarily by the pump laser linewidth, which in our case is $<$ 100 kHz. The instrumental line-shape is negligible, since narrow comb lines sample the molecular spectrum and the FTIR simply acts as a detector array for each of the measured comb lines. In our case, the FTIR is operated with a low resolution of 15 GHz, which enables fast acquisition. The pump (and comb) are tuned over 16 GHz (0.53 cm$^{-1}$), which corresponds to a 100-mA change in the TEC current. By monitoring the beatnote between the pump laser and a femtosecond laser, the pump frequency can be precisely tuned by controlling the piezo voltage. Figure 3a shows the recorded FTIR spectra, where the pump laser is tuned with 80-MHz steps, within the acetylene fingerprinting region. The spectra of selected four comb lines are shown in Fig. 3b where the first three comb lines are clearly overlap with molecular absorption features while the fourth comb line is not. The frequency accuracy of the measurement is equal to the long-term stability of the pump laser which is less than 10 MHz. Since the comb spacing is dependent on the chip temperature, the frequency shift of the pump mode is slightly different from that of the rest of the comb lines. Assuming the pump frequency $f_p = f_{ceo} + mf_r$, the index change as the pump is tuned by $\delta f_p$ is $\delta n = -\frac{m*c}{L(f_p)^2}*\delta f_{p}$ where \emph{c} and \emph{L} are the speed of light in vacuum and the cavity length, respectively. The change in the repetition rate can be expressed as $\delta f_r = -\frac{L(f_r)^2}{c}*\delta n = m{\frac{f_r}{f_p}}^2*\delta f_p$; thus, $\delta f_r = \delta f_p/m$. For the comb line ($n$ = $N$), $\delta f_{n=N}  = (1+\frac{N}{m})\delta f_p$, where the pump mode correspond to $n = 0$ and the higher frequency comb mode has $n > 0$. In our case, $m = 825$ and $f_r$ changes from 127.23 GHz to 127.25 GHz over the 16-GHz tuning range of the pump. Each comb line undergoes a different frequency shift, which equals to the product of its scaling factor and the pump shift. It must be taken into account for the frequency calibration through the scanning process. Figure 4a shows the calculated transmittance of acetylene over the region of 3 - 3.1 $\mu$m where absorption features are clearly observed and their line parameters are retrieved.  The scanning process is completed in one minute which is largely limited by the thermal response time of the microresonator-TEC system. The use of heaters integrated with the microresonator allows for a much shorter response time of sub-ms level\cite{joshi_thermally_2016}, which can be used to significantly increase the scanning speed. Other factors such as the laser tuning speed, the quality factor of the microresonator, and the FTIR acquisition time should also be considered. Figure 4b shows that our transmittance measurement successfully captures different absorption lineshapes. The transmittance plot is compared to the computed transmittance profile using the HITRAN database with Lorentzian lineshapes based on the fact that the gas cell is filled with 40 Torr of acetylene and 415 Torr of nitrogen. Our scheme can also be applied to Doppler-broadened linewidths. Figure 5a plots one of the measured absorption features fitted with a Lorentzian profile where a full-width-half-maximum linewidth of 6.4 GHz is measured centered at 3060.3 nm. The fit residuals are shown in Fig. 5b and has a standard deviation of $4 \times 10^{-3}$. 

\begin{figure}[t]
\centering
\centerline{\includegraphics[width=5.5cm]{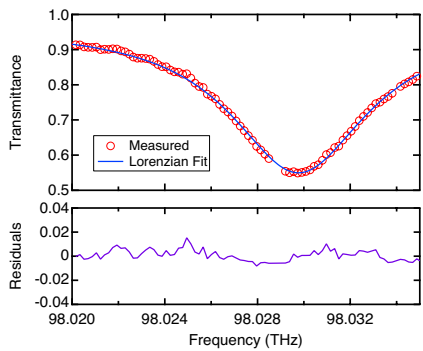}}
\caption{Line parameter measurements. A Lorenzian profile (blue) fits the experimental measurement (red dots). A full-linewidth of 6.4 GHz is measured centered at 3060.3 nm. The residuals (purple) do not show any systematic deviation, and the standard deviation is $4 \times 10^{-3}$.}
\label{Fig5}
\end{figure}

In summary, we demonstrate a highly promising approach for a high-spectral-resolution spectroscopy using a scanning microresonator-based OFC with a mode spacing of over 100 GHz. The ability to scan a broadband OFC, equivalent to tuning a massive array of single-frequency lasers, enables the high spectral resolution comparable to single-frequency laser spectroscopy and allows for Doppler-limited spectroscopy. Rapid thermal tuning of the cavity resonance via an integrated heater and a flatter pump power would enable a complete scan over one FSR to access the entire optical spectral range at video rates. The frequency accuracy would be further improved via a stabilized OFC\cite{delhaye_full_2008, savchenkov_stabilization_2013, papp_microresonator_2014, delhaye_phase-coherent_2016}, for example, by self-referencing\cite{delhaye_phase-coherent_2016} or locking to atomic transitions\cite{papp_microresonator_2014, savchenkov_stabilization_2013}.  In the mid-IR, commercial QCL's have shown a broadband tuning flexibility over 60 cm$^{-1}$ and the potential for a fully chip-scale integration with microresonators. With continued improvement in the linewidth reduction of QCL's, our technique would provide an extremely compact, integrated spectrometer in the mid-IR with high sensitivity and high resolution for on-chip gas-phase spectroscopy. 

\textbf{Funding.} Defense Advanced Research Projects Agency (DARPA) (W31P4Q-15-1-0015); Air Force Office of Scientific Research (AFOSR) (FA9550-15-1-0303); National Science Foundation (NSF) (ECS-0335765).
\\\\
\\
\textbf{Acknowledgment.} This work was performed in part at the Cornell Nano-Scale Facility, a member of the National Nanotechnology Infrastructure Network, which is supported by the NSF.

\bibliography{YoungMidIR}


\end{document}